\documentclass[amsmath,twocolumn,showpacs,amssymb,aps,nofootinbib]{revtex4-1}
\usepackage{geometry}
\usepackage{graphicx}
\usepackage{dcolumn}
\usepackage{bm}
\usepackage{epsfig,amsmath}
\usepackage{amssymb}
\usepackage{subfigure,esint}
\usepackage{float}
\usepackage{comment}
\usepackage[usenames,dvipsnames]{color}
\usepackage[pagebackref=false, colorlinks=true]{hyperref}
\definecolor{redish}{rgb}{0.7,0.2,0.0}  
\definecolor{bluish}{rgb}{0.2,0.5,0.8}

\hypersetup{linkcolor=redish,          
	citecolor=magenta,        
	filecolor=magenta,      
	urlcolor=red}          

\begin{document}
\author{Mamta Gautam}\email{mamtag@iitk.ac.in}
\affiliation{
Department of Physics, Indian Institute of Technology Kanpur, \\ Kanpur 208016, India}
\author{Kunal Pal}\email{kunalpal@iitk.ac.in}
\affiliation{
Department of Physics, Indian Institute of Technology Kanpur, \\ Kanpur 208016, India}
\author{Kuntal Pal}\email{kuntal@iitk.ac.in}
\affiliation{
Department of Physics, Indian Institute of Technology Kanpur, \\ Kanpur 208016, India}
\author{Ankit Gill}\email{ankitgill20@iitk.ac.in}
\affiliation{
Department of Physics, Indian Institute of Technology Kanpur, \\ Kanpur 208016, India}
\author{Nitesh Jaiswal}\email{nitesh@iitk.ac.in}
\affiliation{
Department of Physics, Indian Institute of Technology Kanpur, \\ Kanpur 208016, India}
\author{Tapobrata Sarkar}\email{tapo@iitk.ac.in}
\affiliation{
	Department of Physics, Indian Institute of Technology Kanpur, \\ Kanpur 208016, India}
\title{Spread complexity evolution in quenched interacting quantum systems}

\begin{abstract}
We analyse time evolution of spread complexity (SC) in an isolated interacting quantum many-body system when 
it is subjected to a sudden quench. The differences in characteristics of the time evolution of the SC 
for different time scales is analysed, both in integrable and chaotic models. For a short time after the quench, the
SC shows universal quadratic growth, irrespective of the initial state or the nature of the Hamiltonian, with  
the time scale of this growth being determined by the local density of states. The characteristics of the SC in the next 
phase depend upon the nature of the system, and we show that depending upon whether the survival probability of an initial state
is Gaussian or exponential, the SC can continue to grow quadratically, or it can show linear growth. 
To understand the behaviour of the SC at late times, we consider sudden quenches in two models, a full random 
matrix in the Gaussian orthogonal ensemble, and a spin-$1/2$ system with disorder. We observe that
for the full random matrix model and the chaotic phase of the spin-$1/2$ system, the complexity shows linear growth at early times and
saturation at late times. The full random matrix case shows a peak in the intermediate time region, whereas this feature is less prominent in the spin-$1/2$ system,
as we explain. 
\end{abstract}
\maketitle

\section{Introduction}
The concept of complexity in the context of a quantum mechanical state refers to how `difficult' it is to construct the desired state 
via some pre-assigned basis states and operators. Even though it is a widely used measure in quantum computation research,
the recent flurry of activity in quantum field theory and statistical systems started after the work of \cite{Jefferson:2017sdb}, 
where a proposal for circuit complexity of a quantum state was described, based on Nielsen's
geometric approach to the circuit complexity \cite{Nielsen1, Nielsen2, Nielsen3}. The Nielsen complexity and various other 
related notions of complexity of a quantum state under a unitary evolution was proposed and studied subsequently in several 
works \cite{Khan:2018rzm, Hackl:2018ptj, Bhattacharyya:2018bbv} (see the review \cite{Chapman:2021jbh} for a compendium of
related works). 

Besides these measures of circuit complexity, in a slightly different context of probing operator growth in quantum-many body 
systems, the notion of Krylov complexity (KC) was proposed in \cite{Parker:2018yvk}. The central result of this work was 
embodied in the `universal operator growth' hypothesis, which states that for a non-integrable quantum many-body system in 
the thermodynamic limit, the so-called Lanczos coefficients must grow linearly with $n$
 (with logarithmic corrections present  
for one dimensional systems, where $n$ denotes the index of the ordered Krylov basis). This implies that the corresponding operator complexity must grow exponentially fast, a feature quite 
generic in quantum chaotic systems. Since then, KC has become a popular measure to study various aspects of 
quantum many-body dynamics in and out of equilibrium. For a partial set of works see \cite{Barbon:2019wsy, Bhattacharjee:2022vlt} 
for the use of Krylov complexity in operator growth, \cite{Dymarsky:2019elm, Dymarsky:2021bjq, Kundu:2023hbk} for works in CFTs, 
\cite{Bhattacharya:2022gbz, Bhattacharjee:2022lzy, Bhattacharya:2023zqt} for works on open systems, \cite{Bhattacharyya:2023dhp} for 
KC in Bosonic systems, \cite{Kim:2021okd} as a tool of probing 
delocalization properties in integrable quantum systems, \cite{Hashimoto:2023swv, Camargo:2023eev} focuses on billiard systems. 
Important steps has also been taken to understand features of KC in  QFTs \cite{Avdoshkin:2022xuw, Camargo:2022rnt}. Various other 
important results have also been reported in \cite{Yates:2021asz} - \cite{Erdmenger:2023shk}. 


From a viewpoint more in line with the Nielsen-like complexities, where the complexity is defined as the minimum of a `cost-functional' 
associated with the evolution, the idea of spread complexity (SC) was proposed in \cite{Balasubramanian:2022tpr}. In that work, it was 
shown that the cost that measures the spread of a reference wave function in a fixed basis of states on the Hilbert space under a unitary 
Hamiltonian evolution is the minimum when computed with respect to the  Krylov basis constructed by using the Hermitian operator generating the evolution. 
The associated complexity, the SC, is the analogue of the operator complexity for quantum states.  Starting from 
the work of \cite{Balasubramanian:2022tpr}, the generic features of SC in quantum systems have been explored in a variety of works that 
includes quantum phase transition \cite{Caputa2, Caputa:2022yju, spread1}, work statistics in quantum quench \cite{Pal:2023yik, Gautam:2023pny}, 
probing quantum scar states \cite{Nandy:2023brt}  among others. In this paper our main focus will be SC 
evolution in quantum many-body systems which are far away from equilibrium. 

We note that the evolution of various few-body observables has been studied in the literature,
specifically due to their importance
in quench experiments. However, studies on the quench dynamics of the SC have been relatively rare, and these have been mostly confined to  Hamiltonians
that can be transformed to integrable systems \cite{Caputa2}, \cite{spread1} (there are various works that have studied the evolution 
of other measures of complexities after a quantum quench, see, e.g., \cite{Alves} - \cite{Pal:2022rqq}). 
In this context, the importance of introducing interactions in an otherwise isolated quantum system cannot be over-emphasized,
since integrable systems constitute a small subset of realistic quantum systems with interactions. With this motivation, 
in this work, we go beyond the realm of integrable models, and understand the evolution of SC after a sudden quantum quench is performed in generic
{\it interacting} many-body quantum systems. Here, we highlight the differences in the characteristics of the time evolution shown 
by the SC for different time scales after the quench, both in integrable and chaotic models.
It is known that the exact nature of the dynamics of an interacting quantum system after a sudden quantum quench depends on  
how the local density of states (LDOS) is filled up after the quench. Therefore, the spread of an initial state before the quench in the 
Krylov basis, and hence the SC should crucially depend on how the LDOS behave in that interacting system.  
Our aim in this paper is to find out this connection. We shall highlight how the filling of the LDOS affect the evolution
of the SC after the quench.

To begin with, the basic approach of obtaining the Lanczos coefficients (LCs) and the associated SC after a sudden quantum quench 
used in this paper is the following.
We recall that all information about the LCs, which are the main ingredients of constructing the Krylov basis by means 
of the Lanczos algorithm \cite{Viswanath, Lanczos:1950zz, Parker:2018yvk, Balasubramanian:2022tpr}, are encoded in the moments of the so called 
auto-correlation function, which measures the overlap between the evolved quantum state with the initial one.
Once this is known, 
the full set of Krylov basis can be constructed from the moments of the auto-correlation function at $t=0$, where $t$ denotes the time 
after the quench \cite{Viswanath}.
Typically, for the Hamiltonian evolution, there are two sets of LC, which are denoted by $a_n$ and $b_n$. The first 
set of coefficients give the expectation value of the Hamiltonian in each Krylov basis, and the second set  
represents  the normalization constants for these bases.
Therefore, we can extract all the relevant information about the dynamics in Krylov space, 
and the time evolution of SC, once we have an analytical (or numerical) expression for the auto-correlation function in hand. In this context, 
we note that the characteristic function (CF, which is just the complex conjugate 
of the auto-correlation function) is one one of the most commonly studied quantities in sudden quenches
of interacting quantum many-body system and by itself can reveal important information about the 
nature of the post-quench Hamiltonian and the initial state before the quench. For example, usually when the
CF after a quench decays exponentially, it means that the post-quench Hamiltonian is chaotic in nature \cite{Peres, Cerrtui}, 
however, whenever the initial state is `sufficiently  delocalised' in the energy
basis, similar exponential decay can also be observed in integrable systems \cite{Emerson}.

With this discussion in mind, in this work we first explore in detail the evolution of SC in  generic interacting lattice quantum-many body 
systems, where the analytical form of the auto-correlation functions are known in the literature for a wide range of such systems
 at different time scales after a sudden quench. In many cases approximate 
 analytical expressions for such  auto-correlation functions have been obtained
 by comparing with results from numerical simulations. In this work we  use these generic expressions for such auto-correlation
 functions to obtain the nature of SC evolution for a wide class of realistic interacting quantum many-body systems 
 after a sudden quench.
 
 
In the next section, first we use the expression for the  survival probability (SP) of an initial state before the quench, 
available in the literature for sudden quantum quenches of a parameter of the Hamiltonian of a generic quantum system, 
to comment on the universal features of SC. We find that the SC shows quadratic growth at early times, where the rate 
of the growth is set  by the variance of the local density of states, or equivalently, the LC $b_{1}$. It is to be noted that, 
the quadratic early time behaviour of SC was also reported in \cite{Balasubramanian:2022tpr} for evolution with random matrices. 
Our result points toward the universality of this feature in an interacting quantum many-body system. However, our 
results also suggest that, besides this universal early time quadratic growth, this type of  growth can persist for even larger times 
scales in quenches of  quantum many-body systems 
where the interaction is strong. Next, to explore the behaviour of complexity at late times, we use an interpolating functional form 
of the SP that incorporates between the quadratic decay at early times and exponential decays at late times, valid  
when the external perturbation is not strong, to show that in such a case
the LCs $b_n$ follow a liner growth for relatively small values of $n$ (which determine the early time evolution 
of the complexity) and the SC, after the initial universal quadratic growth, merges into a linear growth at late times.

In the rest of section \ref{early_and_FRM}, we briefly describe the results for  LCs and the SC evolution when one
 uses a  full random matrix (FRM) to describe the Hamiltonian 
of the interacting quantum many-body system
after quantum quench by sampling its elements from a Gaussian orthogonal ensemble (GOE). Here, we use the 
tri-diagonal Hessenberg form of the Hamiltonian to extract the corresponding LCs and later use this form  to find the nature of the SC 
at various time scales. In this context, we note that, in the original work, which introduced the notion of SC \cite{Balasubramanian:2022tpr}, the 
evolution of the SC was explored in detail for quantum chaotic systems modeled by random matrices. It was established that the SC shows 
a characteristic structure - \textit{linear ramp, peak, slope and plateau around a constant value}. These characteristics were further 
studied in \cite{Erdmenger:2023shk}, where it was shown that the late-time features of SC are determined by the probability amplitude 
of each Krylov basis, and is a universal indicator of quantum chaos. In this work we use an analytical form 
for the auto-correlation function after a sudden quench (obtained in \cite{Herrera9}) to  find out an approximate  shape of the 
distribution of $b_n$s.

In section \ref{spin_1/2}, we consider a more realistic model of an interacting quantum 
many-body system, i.e., an interacting spin-$1/2$ model with disorder. 
With increasing disorder, the system transforms from an integrable to non-integrable
phase, and finally to an intermediate limit between the chaotic phase and the many-body
localised phase for large disorder values.
Using a phenomenological expression for the SP after  
a sudden quench, we obtain both sets of LCs and hence the SC when the system is in chaotic domain, as well as  in the intermediate phase. 
Our observations show that in the non-integrable phase, the SC shows linear growth at early times and saturation at late times,
however, the peak in the complexity, present in the FRM case, is less pronounced  here. 
The implications of these results and conclusions are discussed in section \ref{summary}.
This paper also contains an Appendix where we briefly review the construction of the Krylov basis
and the definition of the SC. 

\section{Auto-correlation function and survival probability after a sudden quantum quench}
\label{SP}
We assume that a quantum many-body system with an initial Hamiltonian $H_0$ and prepared in the ground state $\big| \psi_0 \big>$ is quenched
at time $t=0$ to a new Hamiltonian $H$, whose eigenvalues and eigenfunctions
are denoted by $E_n$ and $\big| n \big>$, respectively. To find out the behaviour of the time evolution of the
SC, we need to study time scales associated with the CF, which in this case is given by
\begin{equation}\label{chafunction}
\mathcal{G}(t)=\big< \psi_0 \big|\Psi(t) \big>=\sum_{n}\big|C_0^n\big|^2e^{-iE_n t}~,
\end{equation}
where $C_0^n=\big<n \big|\psi(0)  \big>$ represents the overlap between the initial state and the 
eigenstates of the post-quench Hamiltonian ($\big|n \big>$).
As mentioned in the introduction, the auto-correlation function $\mathcal{S}(t)=\mathcal{G}^*(t)$.
By introducing the LDOS
\begin{equation}\label{LDOS}
\rho_0(E)=\sum_{n}\big|C_0^n\big|^2 \delta (E-E_n)~,
\end{equation}
we can write down the CF as the Fourier transform of the LDOS, i.e.
\begin{equation}\label{CF}
\mathcal{G}(t)=\int dE \rho_0(E) e^{-iE t}~.
\end{equation}
Therefore, once the LDOS is known for  an initial state before the quench and the 
final Hamiltonian, the behaviour of the CF, and hence the auto-correlation function is completely fixed.
Here we also note that for a given final Hamiltonian, the time evolution  of the CF
can be different for different choices of the location of the initial state in the energy spectrum 
of the initial Hamiltonian.

Detailed studies of the survival probability $F(t)$ (which is just the modulus squared of the auto-correlation $F(t)=
\big|\mathcal{G}(t)\big|^2$, i.e. the fidelity) under quenches in integrable and chaotic quantum many-body systems  have
been carried out  in a series of works  \cite{herrera1, herrera2,herrera3,Santos1,Tavora1}.\footnote{This
list of references is indeed incomplete, see references and citations of these works for a more complete review
of the literature on this subject.} Here we briefly describe the behaviour of the SP for different time 
scales after the quench, following the above mentioned references.

For a very short time after the quench, the SP shows a universal quadratic decay, 
independent of the final Hamiltonian or the initial state before the quench. To check this, we first notice that the mean and the 
variance of the LDOS are given, respectively, by 
\begin{equation}
E_0=\sum_{n}\big|C_0^n\big|^2E_n~,~~~\text{and}~~\sigma_0^2=\sum_{n}\big|C_0^n\big|^2 \big(E_n-E_0\big)^2=b_1^2~,
\end{equation}
where in the second expression we have used an identification between the variance of the LDOS and the LC $b_1$ \cite{Pal:2023yik}. 
These two quantities are extremely important in determining the nature of the  subsequent 
dynamics after a quench.
Now expanding the expression for the SP, it is easy to see that, for small times $t\ll\sigma_0^{-1}$, it behaves as \cite{Tavora1,Santos1}
\begin{equation}\label{quadratic}
F(t)\approx 1-\sigma_0^2 t^2 = 1-b_1^2 t^2~.
\end{equation}

After this initial quadratic decay, the behaviour of the SP depends on the strength of the external perturbation (see below). For example, in
systems with two-body interactions, the shape of  the LDOS, as well as the density of states (DOS) is Gaussian, so that 
the resulting shape of the SP is given by the Gaussian function of the form $F(t)=\exp (-\sigma_0 t^2)$ \cite{Flambaum, Izrailev,herrera1}, and this
decay can go up to a saturation. Furthermore, the LDOS can also be of the Lorentzian shape, so that the SP decays exponentially with time.  These  behaviour usually hold  upto a time scale of $\sigma_0^{-1} \lesssim t \lesssim t_P$,
where $t_P$ indicates the time scale where the decay of the SP follows a power law.

At late times therefore,  the SP attains a power law decay of the form $F(t) \propto t^{-\gamma}$, where, the exact value of the exponent
$\gamma$ crucially depends on the nature of the system under consideration, i.e. how the LDOS fills up, as well as  the initial state before the quench. 
A detailed study of this power law decay and calculation of the resulting exponent has been performed in \cite{Tavora1},
where it was shown that from the numerical value of $\gamma$, we can predict whether a given initial state will thermalise or not,
therefore providing a way to probe the thermalisation of an initial state based only on the dynamics after quench. It is thus clear that
the exact nature of the dynamics of the SC will therefore depend on how the LDOS is filled, in the system under consideration.
In this paper, our broad goal is to perform such analysis about the evolution of SC depending upon the different types of LDOS filling after the quench.

\section{Behaviour of spread complexity at early times after a sudden quench}\label{early_and_FRM}

Before considering quench dynamics in some specific models, here we first study the approximate evolution
of the SC at different time scales  after the quench, and the behaviour of the resulting LC  
using the  evolution of the SP as described above. 
We assume that the Hamiltonian of the quantum system under consideration 
can be written as $H=H_0 +g V$, where $H_0$ is the unperturbed
Hamiltonian, and $V$ is an external perturbation, and $g$ denotes the strength of the external perturbation.
We  also assume that the state $\big| \psi_0 \big>$ of the system before the quench is the first state of the Krylov basis 
i.e., $\big| K_0 \big>=\big| \psi_0 \big>$. The construction of the Krylov basis 
and the definition of the SC are briefly described in Appendix \ref{krylov}. All 
other relevant details, which have been used in our numerical analysis below, can be found in 
\cite{Viswanath, Lanczos:1950zz, Balasubramanian:2022tpr}.

\subsection{Analysis for Gaussian exponential decays}
First we consider the time scale $t \ll \sigma_0^{-1}$, where, as we have discussed above, the SP decays quadratically with
time irrespective of the nature of the system  under consideration or the initial state before the quench.
 Since before the quench, the state of the system is the 
lowest state of the Krylov basis, for times scales $t\ll\sigma_0^{-1}$ just after the quench, the time evolved state $\big|\Psi(t) \big>$
will spread over only the first few Krylov basis elements with small basis number $n$. Now we can use the auto-correlation function
of the form $\mathcal{S}(t)=1-\frac{1}{2}\sigma_0^2 t^2$ to obtain the first few LCs as well the SC. It can be checked that
for this auto-correlation function, $a_n=0$, while $b_n$s grow with $n$ and 
are proportional to the width $\sigma_0$ of the LDOS. 
\footnote{The LC and, subsequently the SC are obtained 
	from this SP by using the procedure reviewed in Appendix \ref{krylov}. } 
In Fig. \ref{fig:bn_early}, we have provided the numerical values of the
first few $b_n$s with $\sigma_0=1$.

\begin{figure}[h!]
	\begin{minipage}[b]{0.8\linewidth}
	\centering
		\includegraphics[width=0.85\textwidth]{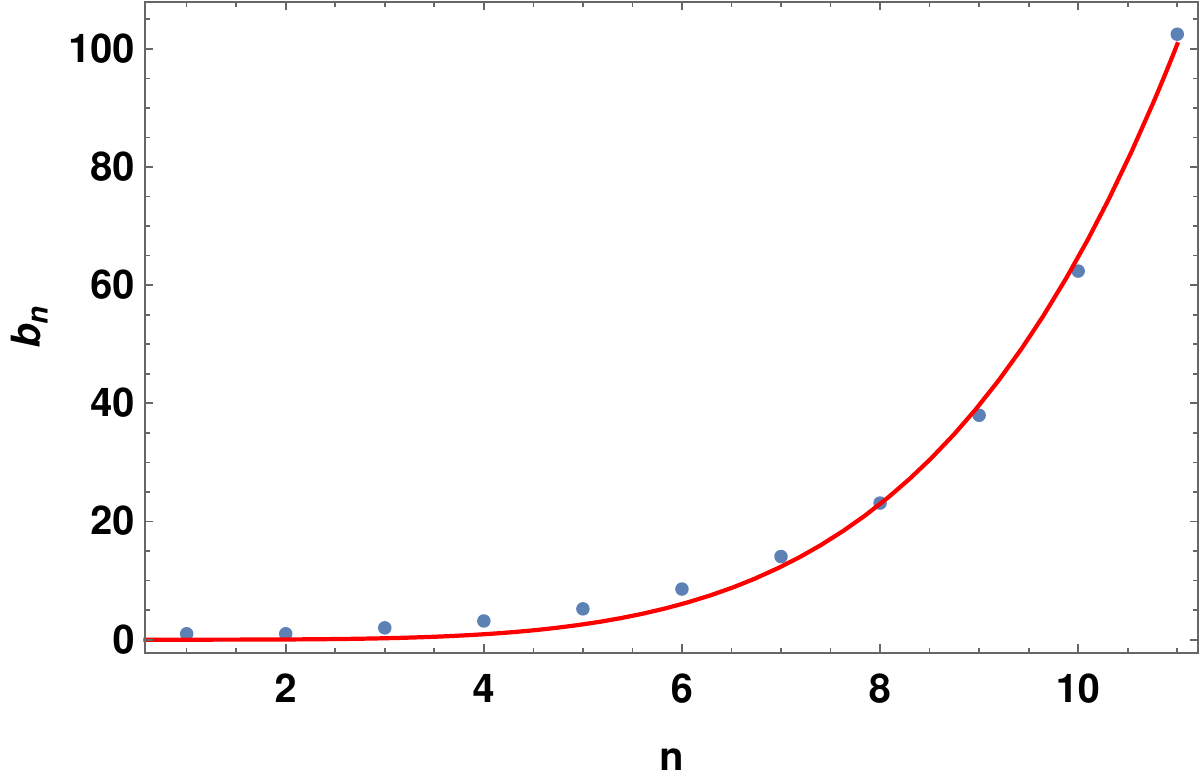}
	\caption{Numerical values of the first few $b_n$s. Here we have set $\sigma_0=1$, and $b_n$s are proportional to
		$\sqrt{\sigma_0}$. The red curve is of the form $n_1 n^{n_2}$, and provides a good fit of the $b_n$s with
		$n_1 \approx 0.0014$ and $n_2 \approx 4.6446$. }
	\label{fig:bn_early}
		\end{minipage}
   \end{figure}
    \begin{figure}
    \begin{minipage}[b]{0.8\linewidth}
    		\centering
    	\includegraphics[width=0.85\textwidth]{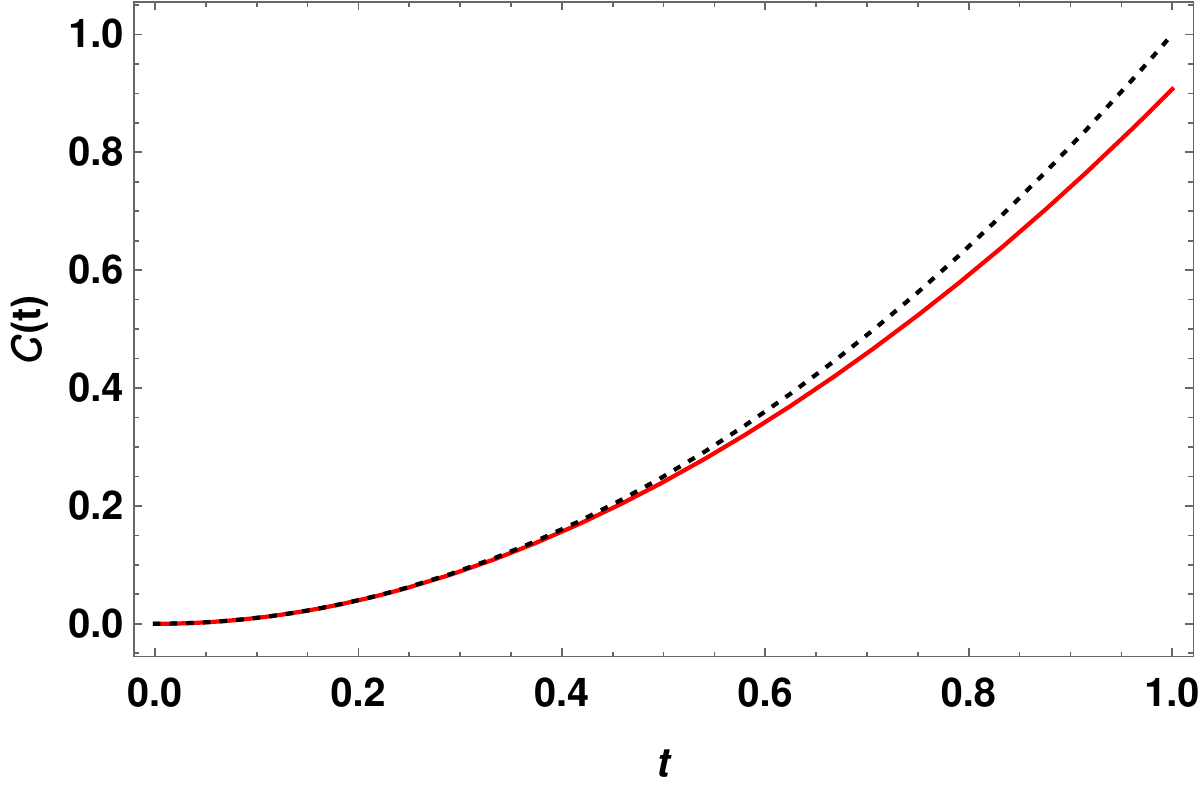}
    \caption{Time evolution of SC (red curve) for $t\ll\sigma_0^{-1}$ after the quench. Here we have set $\sigma_0=b_1=1$. The
    	dotted black curve is the plot of $t^2$, which have a reasonably accurate matching with complexity at early times. }
    \label{fig:SC_early}
    \end{minipage}
\end{figure}

The time evolution of the SC  for this case is shown in  Fig. \ref{fig:SC_early}.  Along with the complexity, calculated numerically (the 
red curve), we have also shown the plot of $t^2$, which reasonably approximates the SC curve at early times. Since the quadratic decay
of the SP after quench is universal, we conclude that irrespective of the nature of the quantum many-body system under consideration,
and the initial state before the quench, the SC will always grow quadratically, with the coefficient of the growth being determined
by the variance of the LDOS, or equivalently, the LC $b_1$. 

As we have discussed in the Introduction, after this initial quadratic decay, the nature of the time evolution of the SP and the auto-correlation 
function will depend on the exact nature of the quantum system under consideration. Two of the most common types of
time evolutions encountered in the literature  for quenches in interacting many-body quantum systems are the Gaussian and the exponential decays. These two types of decays appear
in  a chaotic system  precisely 
when the  shape of the LDOS is Gaussian and a Breit-Wigner form, respectively.

In the presence of strong interactions,\footnote{See \cite{Flambaum, Izrailev} for the exact quantification of the interaction
	strength in the context of two-body random interaction models.} the shape of the LDOS is Gaussian, so that the SP  can be a function of the variable $\sigma_0^2 t^2$
for a time scale much  longer than $t_0<< \sigma_0^{-1}$ \cite{Flambaum2}, and the auto-correlation function 
(obtained from Eq. \eqref{chafunction}) is of the form 
$S(t)\approx \exp \big(-\frac{\sigma_0^2}{2} t^2\big)$. 
For such a Gaussian form for the auto-correlation, the behaviour for the LC and the SC are well known,
see e.g., \cite{Caputa1,Balasubramanian:2022tpr}. Here the LCs are given by $a_n=0$, and $b_n=\sigma_0 
\sqrt{n}$, and the resultant SC grows quadratically with time, with the coefficient of this
growth being determined by the variance $\sigma_0$  of the LDOS. Therefore, in the presence of strong interactions,
the initial quadratic growth of the SC persists for time scales longer than $t_0(\ll \sigma_0^{-1})$.

 In many cases, even in the presence of Gaussian LDOS,
an initial Gaussian decay can change to an exponential decay before reaching saturation 
\cite{Flambaum, Flambaum2} (see below for dynamics of the SC in such cases). However, 
the Gaussian decay can also persist until saturation. This was indeed shown to be the case for sudden quenches in XX model, 
and spin-$1/2$ systems with impurities in  \cite{Herrera7}. Therefore, for quenches in these systems, the characteristic 
quadratic growth of the SC continues upto the saturation point of the SP.

For perturbations that are not very strong ($g<1$), the long time decay of the SP is exponential, and hence  it can be written as 
$F(t)\approx \exp \big(-\Gamma t\big)$, where $\Gamma$ is the width of the 
corresponding Breit-Wigner distribution of the LDOS.  
In this case, it is interesting to consider an extrapolation formula for the SP derived in  \cite{Flambaum2},
which interpolates between the initial quadratic and the long time exponential decays. The proposed form for the 
CF can be written as 
\begin{equation}\label{extrapolation}
\mathcal{G}(t)=\exp \Bigg[\frac{\Gamma^2}{4 \sigma_0^2}-\frac{1}{2}\sqrt{\frac{\Gamma^4}{4 \sigma_0^4}+\Gamma^2 t^2}\Bigg]~.
\end{equation}
It can be easily checked from this expression that for $\Gamma < \sigma_0$, this functional form for the CF represents the 
initial quadratic decay of the form in Eq. \eqref{quadratic}, while at late times it gives rise to $F(t) \approx \exp \big(-\Gamma t\big)$. Furthermore, we also note that for this form 
for the SP, the width (i.e. the second moment) of the corresponding LDOS is not 
$\sigma_0$.
We use this CF to calculate the LC and the time dependence of the SC. 

\begin{figure}[h!]
	\begin{minipage}[b]{0.85\linewidth}
		\centering
		\includegraphics[width=0.8\textwidth]{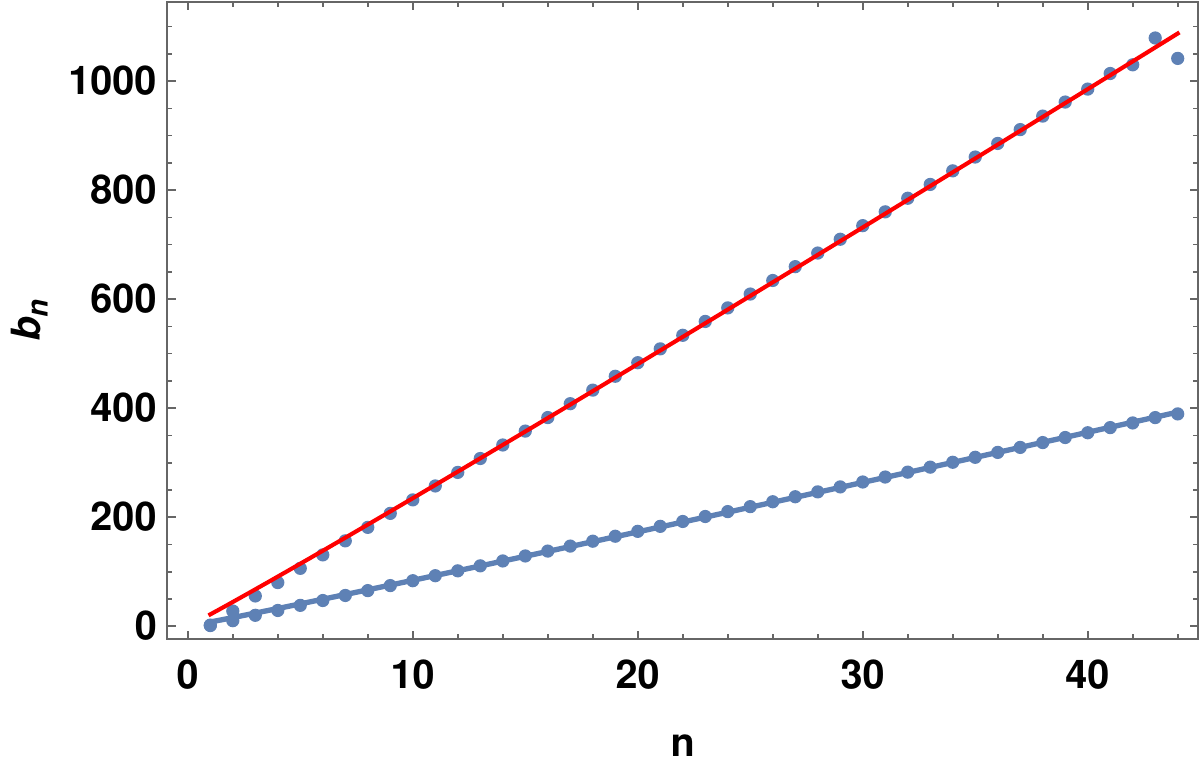}
		\caption{Numerical values of the first few $b_n$s calculated from the CF in Eq. \eqref{extrapolation}.
		 The red dots  are with $\sigma_0=2$, and blue dots are with $\sigma_0=1.2$, where in both the cases 
	 $\Gamma$ is fixed to $0.5$. The straight lines show the linear fitting to the numerically evaluated
 $b_n$s. }
		\label{fig:bn_extp}
	\end{minipage}
\end{figure}

The first set of LC $a_n$ calculated from the CF in Eq. \eqref{extrapolation} are
zero in this case as well, while the first few $b_n$s are shown in Fig. \ref{fig:bn_extp}
for two different values of the 
variance of the LDOS.  As can be seen in both cases, these $b_n$s evaluated numerically can be fitted with 
curves of the form $b_n=n_1 n^{n_2}$, where the exponent $n_2$ is approximately equal to unity in both the 
cases, i.e. $b_n \approx n_1 n$. The slope of the straight line fitting depends on the variance $\sigma_0$ 
in such a way that  higher the variance, the higher is the slope of the straight-line fitting. For example, for the red dots with
$\sigma_0=2$ we find $n_1=21.687$, and for the blue dots, plotted with $\sigma_0=1.2$ we obtain $n_1=7.682$.
Therefore, for systems with  LDOS of  same shape but different widths, the growth rate of $b_n$s is higher when width of the LDOS is greater, which in turn results in faster decay of the SP.

\begin{figure}
	\begin{minipage}[b]{0.85\linewidth}
		\centering
		\includegraphics[width=0.8\textwidth]{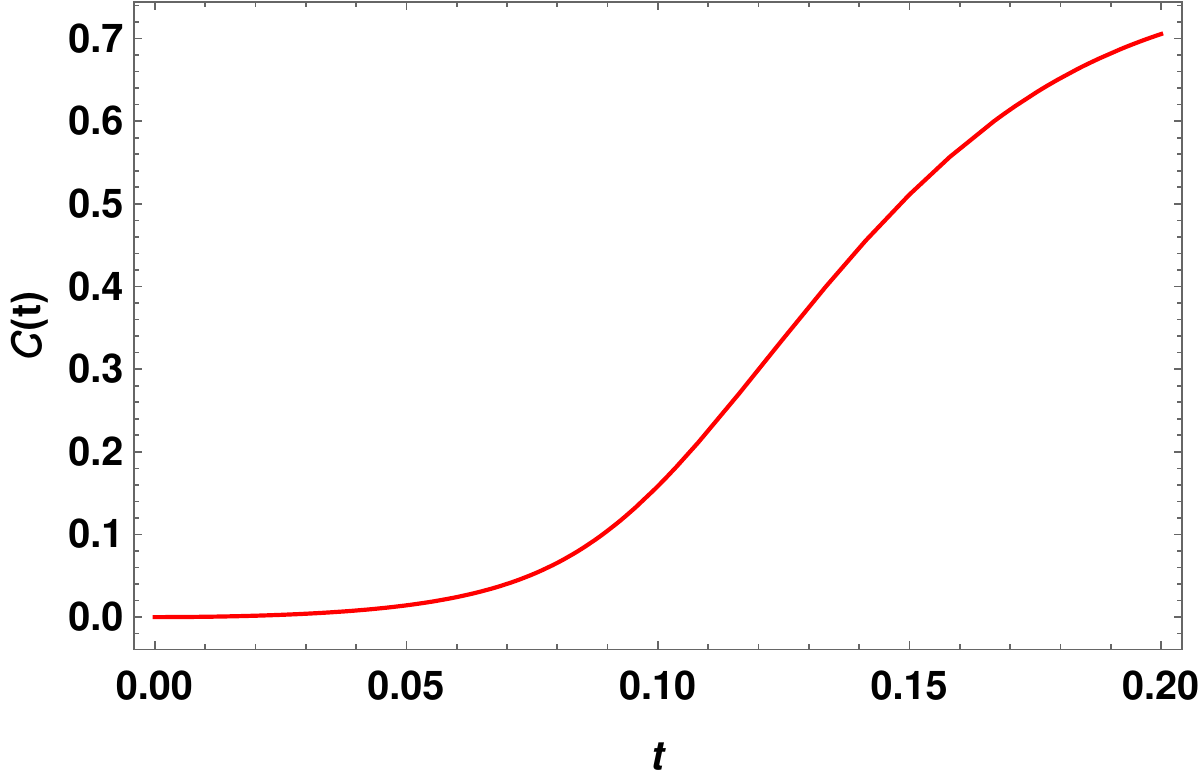}
		\caption{Time evolution of SC with $\sigma_0 =2$ and $\Gamma=0.5$ (red dots in Fig. \ref{fig:SC_extp}).
		The initial quadratic growth transforms to a linear growth at late times, with the time scale of the 
	transformation being determined by the variance of the LDOS. }
		\label{fig:SC_extp}
	\end{minipage}
\end{figure}

The time evolution of the SC (with  values of $\sigma_0 =2$ and $\Gamma=0.5$, for the red dots in Fig. \ref{fig:bn_extp})
is shown in Fig. \ref{fig:SC_extp}. For other values of the parameters, the features shown by the time evolution of the 
SC are similar. As can be seen,  the initial quadratic growth of the SC (valid for the time scale $t\ll\sigma_0^{-1}$)
transforms to an approximately linear growth at late times.

Since the variance of the LDOS $\sigma_0$ is the quantity which sets the time scale of the initial quadratic growth, 
the time where the quadratic growth matches with the 
linear growth occurs at a later time in the second case (blue dots) considered in Fig. 
\ref{fig:bn_extp}, than the one first case (red dots). Therefore, when $\sigma_0^{-1}=1.2$,
the quadratic growth of the SC persists for a longer amount of time, than when $\sigma_0^{-1}=2$.
In the opposite case, where we fix the variance of the LDOS, and change $\Gamma$, the higher 
value of $\Gamma$ leads to lower values of the slope of the linear fit for the $b_n$s.
This observation clearly illustrate the role of the initial state before the quench on the growth rate of the 
	corresponding LCs, and hence the SC.

\subsection{Analysis with full random matrices}
After finding out the behaviour of the SC at time scales just after the quench ($t_0<< \sigma_0^{-1}$), and
in intermediate times before the power law of the SP sets up for a wide class of generic
interacting quantum many-body system, in this  section we consider evolution of SC late times after quench.
First we consider time evolution when the post-quench Hamiltonian is modeled by full random matrices.

Indeed, one of the most common approaches used to study a strongly chaotic quantum system is to model
them as full random matrices (FRM). A quantum chaotic system usually shows Wigner-Dyson
distribution of the spacings of neighboring energy levels due to the presence of strong  
level repulsion \cite{Mehta,Guhrrev}. 
In this subsection, we assume the quenched Hamiltonian to  be a
FRM from the Gaussian orthogonal ensembles (GOE), so that it is actually possible to obtain an
analytical expression for the SP and hence the CF \cite{Santos1}. Modeling a quantum 
many-body system as a FRM 
is clearly not a realistic choice, since it implies simultaneous as well as  infinite-range interactions between  all the particles of the system. However, in this case,  we can use the 
analytical formula for the CF to gain insights into the nature of the LC and the complexity 
evolution after quench to apply these in a more realistic model considered in the next section. 

In this context, we mention that the LC and the SC in evolution with random matrices (RM) have been
studied in details in \cite{Balasubramanian:2022tpr}, where  universal characteristics of the SC evolution
for such RM models were established.\footnote{See also, \cite{Barbon:2019wsy, Rabinovici:2021qqt, Kar:2021nbm} for some related works on SC in the contexts of the RM. } 
In this section, our main goal is to use the analytical expression for the SP 
obtained when  FRM is used to model the Hamiltonian after a quench \cite{Santos1, Herrera4}, and to
understand the  effect of two-level form factor (which is non-vanishing only
in  systems that have correlations between energy levels) in the expressions for the LC and the time evolution of the  SC.
 
As before, we assume that the Hamiltonian after a quench is given by $H=H_0 +V$, where $H_0$ is the unperturbed
Hamiltonian, and $V$ is an external perturbation, i.e.,  by quenching we give a non-zero value to it.\footnote{Here we have set the strength of the perturbation to $1$ for convenience.}  
When we use the FRM, the matrix elements of the Hamiltonian $H_{nm}$ are random numbers, and since we are considering
the GOE, these numbers are taken from a Gaussian distribution with a mean value of zero.
The expression for the SP for a quenched quantum system modeled by FRM belonging to the GOE can be written as \cite{Herrera9, Santos1}
\begin{equation}\label{SP_FRM}
\big<F(t)\big>_{FRM}=\frac{1-\big<\bar{F}\big>_{FRM}}{\mathcal{N}-1}\bigg[4\mathcal{N} \frac{J_1^2(\eta t)}{(\eta t)^2}-\mathcal{B}_2 \Big(\frac{\eta t}{4 \mathcal{N}}\Big)\bigg]+\big<\bar{F}\big>_{FRM}~.
\end{equation}
Here, $\big<.\big>_{FRM}$ denotes the ensemble average,  $\mathcal{N}$ is the size of the RM under consideration, 
$\eta=\sqrt{2\mathcal{N}}$
and $J_1(t)$ is the Bessel function of the first kind. Furthermore, $\bar{F}=\sum_{n}\big|C_0^n\big|^4$,
so that for GOE FRM we have $\big<\bar{F}\big>_{FRM}=3/(\mathcal{N}+2)$.
The functional form for the time dependence of the function $\mathcal{B}_2 (t)$, known as the two-level form 
factor  is given by \cite{Herrera9, Santos1}
\begin{equation}
\mathcal{B}_2 (t)=\Big[1-2t+t \ln (1+2t)\Big]\Theta(1-t)+\Big[t\ln \Big(\frac{2t+1}{2t-1}\Big)-1\Big]\Theta(t-1)~,
\end{equation}
where $\Theta$ is the Heaviside step function.

The function $\mathcal{B}_2 (t)$ is non-zero only when the energy levels of the Hamiltonian are correlated. For integrable systems, where the
levels are uncorrelated, the two-level form factor vanishes, i.e., $\mathcal{B}_2 (t)=0$. The quantity  $\big<\bar{F}\big>_{FRM}$ 
determines the long-time saturation value of the SP, so that at very late times after the quench, the SP only
fluctuates around this constant value. The time evolution curve of the SP can be divided into the following three characteristics
regions. (1) The Bessel function term appearing above governed  the  initial decay, which is of the form  $1/t^3$ 
\cite{Tavora1,Tavora2}. (2) At very late times, the SP acquires a saturation value $\big<\bar{F}\big>_{FRM}$. (3) Between 
the power law decay and the final saturation, there is a dip in the SP (below the saturation value) due to the 
presence of the two-level form factor (and hence correlations between the energy levels) which is known as the
correlation hole \cite{Leviandier, Guhr, Alhassid}. When the quantum system under consideration is an integrable one,
this  correlation hole region vanishes from the dynamics of the SP.

From the above discussion, it is clear that if we calculate the LCs and the resultant SC
using the SP given in Eq. \eqref{SP_FRM} for two different cases, first with the full analytical expression given
in this equation valid for FRM in the GOE, and the second one with $\mathcal{B}_2(t)=0$, we expect that the LC and the SC
will show universal characteristics features of a quantum chaotic system, established in \cite{Balasubramanian:2022tpr} for the first case only.

First, we consider the case when the two-level form factor is zero. In this case, we obtain the CF
by directly using Eq.  \eqref{CF}, and the fact that here the ensemble average $\big<\delta (E-E_\alpha)\big>_{FRM}$
is just the density of states which we denote as $R(E)$. Now using the well-known fact that  for 
FRM, the LDOS and the DOS are equal, and both have a semi-circular form \cite{Guhrrev, herrera1, Herrera6}, we obtain that
\begin{equation}
\big<\mathcal{S}(t)\big>_{FRM}=\frac{J_1(2\alpha t)}{\alpha t}~,
\end{equation}
where, $4\alpha$ is the length of the spectrum.
Notice that from this auto-correlation function in the limit of large times, the saturation value of the SP 
is zero, instead of $\big<\bar{F}\big>_{FRM}$.
We can use this auto-correlation function to calculate the LC and the SC of evolution after the quench.


Using the Lanczos algorithm we obtain that here, the $a_n=0$ and $b_n=\alpha$. The SC shows 
linear growth with time after the quench (this linear growth actually comes after a quadratic growth at very early times, 
see the discussion at the beginning of this section). 
Here we also note that,
for a realistic few-body quantum system, considered in the next section, the DOS, unlike the FRM considered here,
is not of semi-circular shape; rather, it  has a Gaussian form. Therefore, the LDOS
of these realistic interacting systems can not exceed the Gaussian shape.

\begin{figure}[h!]
	\begin{minipage}[b]{0.95\linewidth}
		\centering
		\includegraphics[width=0.85\textwidth]{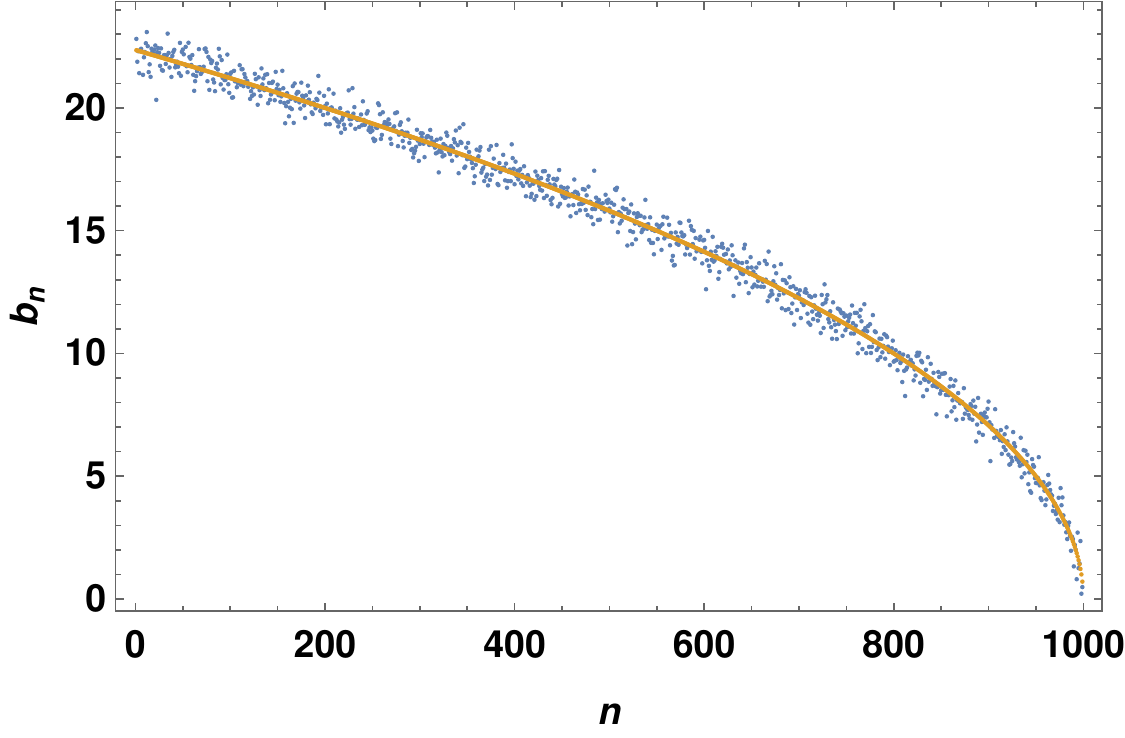}
		\caption{Variation of  $b_n$ with $n$ for the case of FRM in the GOE. Here we have taken $\mathcal{N}=1000$. The solid line represents an approximate fitting for the $b_n$s, and 
		is of the form $b_n=n_1(\mathcal{N}-n)^{n_2}$.
			We fix the unknown constants by using the exact SP in  Eq. \eqref{SP_FRM}.}
		\label{fig:bn_frm}
	\end{minipage}
\end{figure}

Next, we consider the case with non-zero two-level form factor. Here we directly use FRM in the GOE
and obtain the Hessenberg form to find out the LC. From the Hessenberg form  we see that the $a_n \approx0$, and the 
variations  of  $b_n$s with $n$ is shown in Fig. \ref{fig:bn_frm}. The solid line represents an approximate fitting for the $b_n$s, 
and is a curve whose equation is of the form $b_n=n_1(\mathcal{N}-n)^{n_2}$. Below we shall fix the numerical 
values for the two constants appearing in this equation. 
\begin{figure}
	\begin{minipage}[b]{0.95\linewidth}
		\centering
		\includegraphics[width=0.85\textwidth]{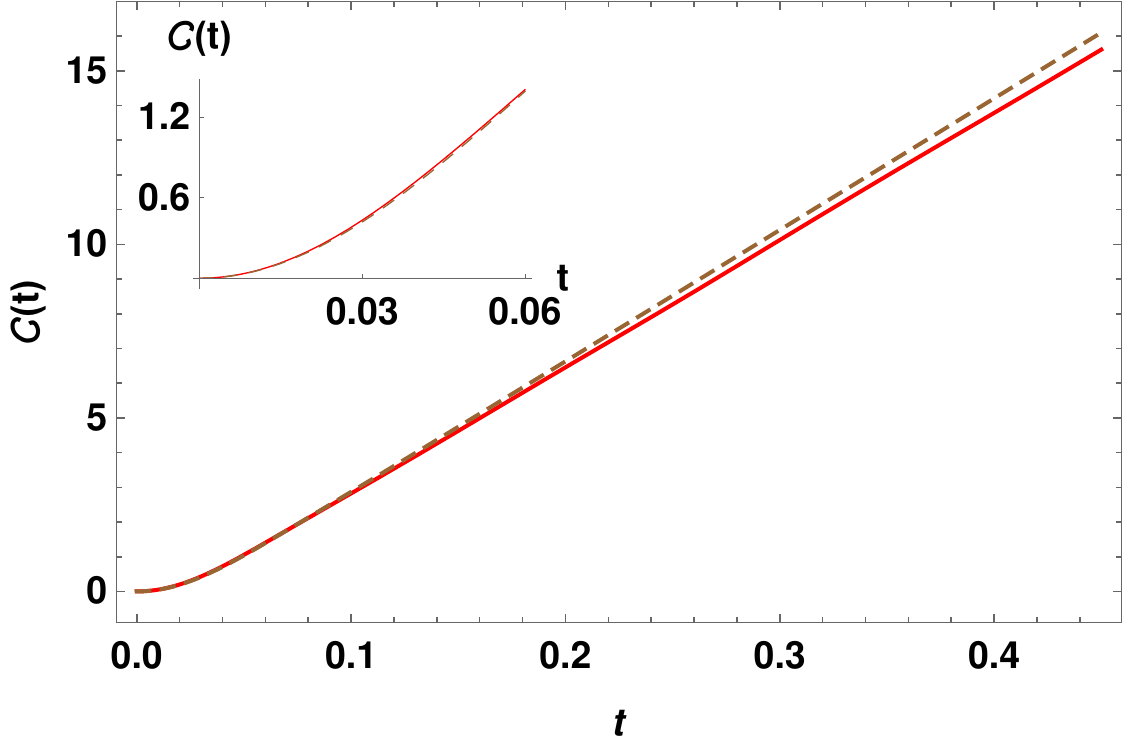}
		\caption{Early time evolution of SC when the quenched Hamiltonian is modeled with FRM in the GOE (the red curve).
			The dashed brown curve represents evolution when the two-level form factor is absent from the SP. In both cases, an early quadratic growth (shown clearly in the inset) merge into a linear growth at late times.}
		\label{fig:SC_frm1}
	\end{minipage}
\end{figure}

In Fig. \ref{fig:SC_frm1} we have shown the early time behaviour of the SC for the FRM case and compared it with the evolution 
in the absence of the  two-level form factor term.  In both cases, the universal quadratic growth at very early times
merged into linear growth at late times. The plots for SC with and without correlations between the energy levels
match with each other for early times.  This is due to the fact that in both cases, early time decay of the 
SP is governed by the Bessel function term (and hence a power law decay), and only after this initial power law decay, the  effect of the 
two-level form factor manifests itself through the presence of the correlation hole in the SP, 
and thus the two curves for the SC  differ from each other only  after this time.

\begin{figure}[h!]
		\centering
		\includegraphics[width=0.45\textwidth]{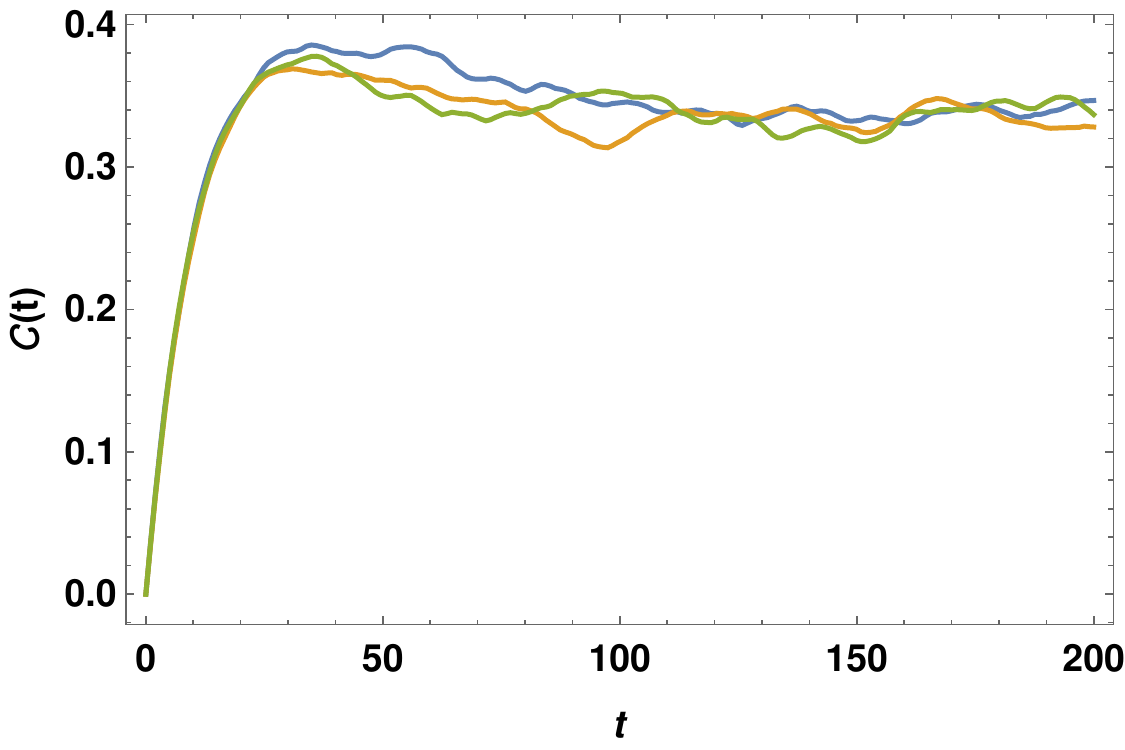}
		\caption{Evolution of the SC after a quench for three different 
			RM in the GOE. Here $\mathcal{N}=1000$. The presence 
		of the correlation hole in the SP is responsible for the peak in the SC.}
		\label{fig:sc_frm2}
\end{figure}

Now to determine the unknown constants in the fitting curve for the $b_n$s we use the exact SP in Eq. \eqref{SP_FRM},
and compute the analytical expressions for the first few $b_n$s. Then we expand these expressions, taking 
$\mathcal{N}\gg1$, and find out the dominant  contribution of $\mathcal{N}$. For example, we have $b_1 \approx 
\sqrt{\frac{\mathcal{N}}{2}}+\sqrt{\frac{1}{2\mathcal{N}}}+\mathcal{O}\big(\frac{1}{\mathcal{N}^{3/2}}\big)$.
Comparing this with the fitting function in the limit $\mathcal{N}\gg1$, we get $n_1=1/\sqrt{2}$ and $n_2=1/2$.
This matching  can be done by taking the analytical expressions for  any higher order $b_n$s as well.
Furthermore, we also notice that, when all the $b_n$s are expanded in powers of $\mathcal{N}$, the leading 
order  contribution which survives in $\mathcal{N} \rightarrow \infty $ limit is proportional to $\sqrt{\mathcal{N}}$.

Time evolution of the SC at late times after quench with different FRM in the GOE is shown in Fig. \ref{fig:sc_frm2}.
In all the cases, after an initial linear growth, the SC reaches a maximum value  and then decays smoothly
to a saturation value at late times. The peak  in the SC for the FRM  is due to the presence of the two-level
form factor (and hence the correlation hole) in the SP. Furthermore, these features are consistent with the universal
nature of the evolution of the  SC of RM models discussed in \cite{Balasubramanian:2022tpr, Erdmenger:2023shk}.

\section{Spread complexity in quenches of interacting spin-$1/2$ models}\label{spin_1/2}

As we have discussed in the beginning of the previous subsection, the FRM is a somewhat unrealistic approximation
of a realistic quantum system. However, the analytical expression for the SP for this FRM models, given in Eq. 
\eqref{SP_FRM}, can be used as a reference to obtain the expression for the SP under quench
for more realistic 1-dimensional (1D) spin-$1/2$ quantum system, which shows chaotic behaviour in certain limits. Using numerical analysis,
it is possible to obtain the SP of an initial state for sudden quench done on such 1D systems. Comparing
these numerical results with the exact analytical expression for the SP for FRM it is possible
to identify the following general dynamical features of quenched 
realistic many-body quantum systems \cite{Herrera6, Santos1}:
(1) An initial  power-law decay. (2) Presence of a correlation hole, where the SP decrease below the saturation value, and (3) 
Saturation at late times.  The saturation value of the SP is equal to its infinite time
average, i.e. $\bar{F}(t)\sim \sum_{n}\big|C_0^n\big|^4$, which, in turn, is just the  inverse of the participation ratio of the initial state.
In this section we demonstrate how these typical features of SP affect the evolution of SC at different times scales after when a sudden  quench is performed in such a spin-$1/2$ system.

The Hamiltonian we consider is that of a one dimensional spin-$1/2$ system, which has two parts,  
$H=H_0+V$, i.e, the original Hamiltonian $H_0$, and a perturbation $V$,
respectively.  Here we assume these to be of the form  \cite{Santos1}
\begin{equation}
H_0=\sum_{j=1}^{L}h_j S_j^z ~,~~V=\sum_{j=1}^{L} \Big(S_j^x S_{j+1}^x
+S_j^y S_{j+1}^y+S_j^z S_{j+1}^z\Big)~,
\end{equation}
where $L$ is the total number sites, and periodic boundary conditions are 
assumed to be applied. Since the parameter $h$ is non-zero, the system is disordered, and $h_j$ are  
random numbers taken from a uniform distribution $[-h,h]$. From the structure 
of the Hamiltonian, we see that before the quench, the coupling strength $g=0$, and after the quench it abruptly
changes to $1$, so that the system is taken far from equilibrium through the quench. 

Changing the disorder strength 
from an initial zero value, the Hamiltonian $H$ typically shows three distinct regions, a sharp
transition from integrable (corresponding to zero disorder strength) to a chaotic domain, which is followed 
by a chaotic region. Finally, as the disorder strength is increased, the system acquires an intermediate
region between chaotic and many-body localised phases \cite{Herrera8}. The level spacing
distribution corresponding to the total Hamiltonian accordingly starts from Poisson, transforms to Wigner-Dyson, and
then again goes back to Poisson as the disorder strength is increased.

In this paper we consider the largest subspace of the total Hilbert space of the 
system which has $\mathcal{S}_z=0$, and has dimension $N=L!/(L/2)!^2$, where 
$\mathcal{S}_z=\sum_kS^{z}_k$ is the total spin along the $z$ direction and is a conserved quantity for the Hamiltonian under consideration.

For the type of  system  under consideration, after a sudden quench, 
an analytical form for 
the  SP can be obtained by comparing results from numerical simulations
and the SP for the FRM given in Eq. \eqref{SP_FRM}, and this is given by \cite{Herrera9}
\begin{equation}\label{SP_12}
\big<F(t)\big>_{h}=\frac{1-\big<\bar{F}\big>_{FRM}}{\mathcal{N}-1}\bigg[\mathcal{N} \frac{g(t)}{g(0)}-\mathcal{B}_2 \Big(\frac{\sigma_0 t}{ \mathcal{N}}\Big)\bigg]+\big<\bar{F}\big>_{h}~,
\end{equation}
where the $\big<\cdots\big>_h$ now represents the disorder average,  and the functional
form for the function $g(t)$ is given by
\begin{equation}
g(t)=e^{-\sigma_0^2 t^2}+\mathcal{A}\frac{1-e^{-\sigma_0^2 t^2}}{\sigma_0^2 t^2}~.
\end{equation}
The constant $\mathcal{A}$ can be obtained by fitting this function with results
for SP obtained from numerical  simulations.


\begin{figure}[h!]
	\begin{minipage}[b]{0.85\linewidth}
		\centering
		\includegraphics[width=0.95\textwidth]{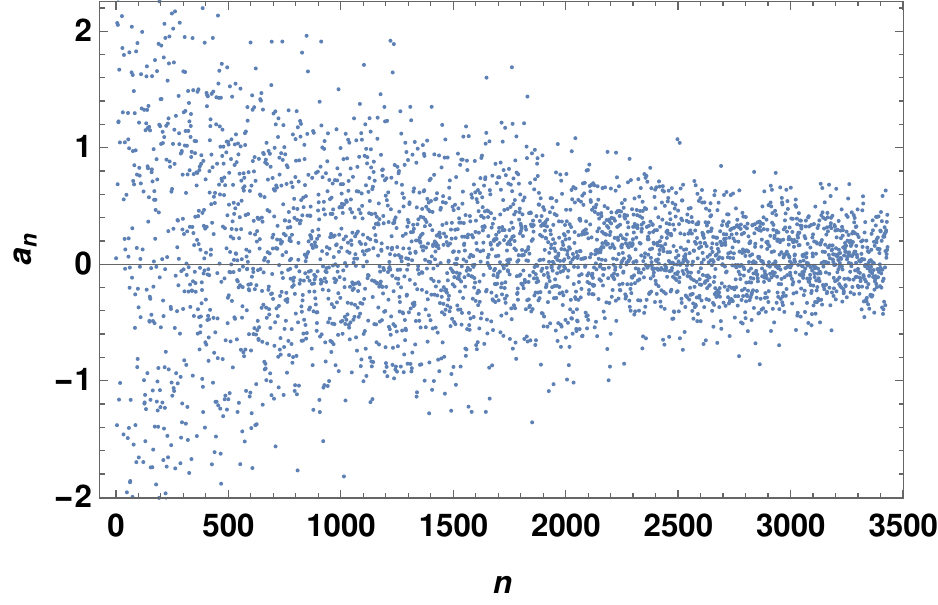}
		\caption{Variation of  $a_n$ with $n$ for the quench in the integrable limit  spin-$1/2$ model with 
			disorder. Here we have taken $L=14$ and the initial state is a domain wall.  }
		\label{fig:an_12int}
	\end{minipage}
\end{figure}
\begin{figure}
	\begin{minipage}[b]{0.85\linewidth}
		\centering
		\includegraphics[width=0.95\textwidth]{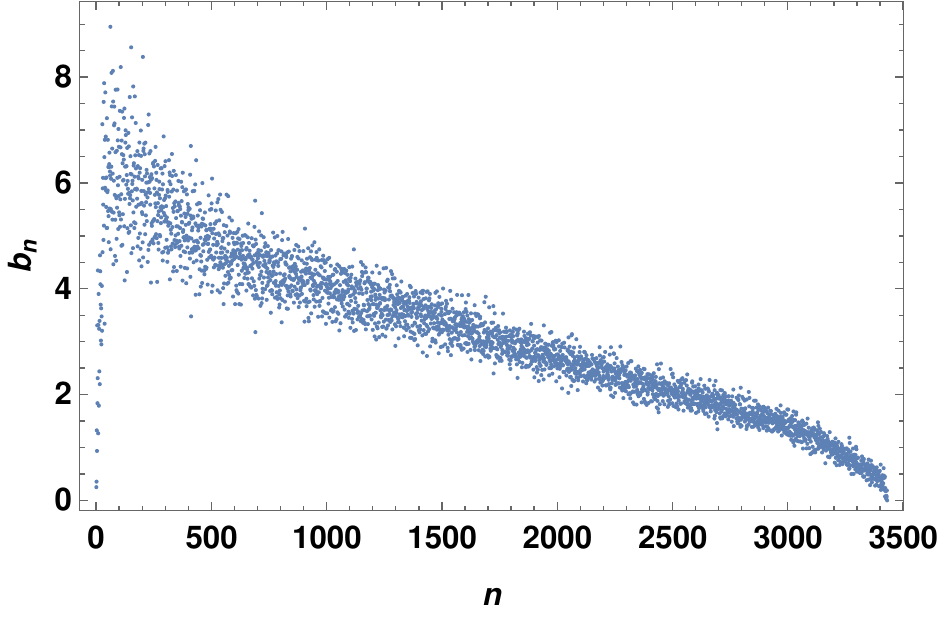}
		\caption{Variation of  $b_n$ with $n$ for the quench in the integrable  spin-$1/2$ model with 
			disorder. Here we have taken $L=14$ and the initial state is a domain wall. }
		\label{fig:bn_12iint}
	\end{minipage}
\end{figure}

First we study the LCs and the SC of the spin-$1/2$ model in the non-chaotic case. 
As the initial state before the quench, we consider a domain wall, i.e.,
$\big| \uparrow \uparrow \uparrow \cdots \downarrow \downarrow \downarrow\big>$.  
The variation of the LCs $a_n$ and $b_n$ with 
respect to $n$, with $L=14$ sites  are shown in Figs. \ref{fig:an_12int} and
\ref{fig:bn_12iint}, respectively. The $a_n$s are distributed almost uniformly around zero.  On the other hand, the $b_n$s show initial sharp growth
for lower values of $n$, reach a maximum peak, then decay gradually towards zero as we reach towards the end of the Krylov basis.
As expected, after initial growth, the SC oscillates with time.

Next we consider quench in the non-integrable limit of this model. Here once again we take the domain wall as the initial state before the quench,
and assume the disorder parameter to be $[-0.4,0.4]$. Here it is easy to see that, compared with the integrable limit of this model, the LCs are less
randomly distributed. In particular, the variation of $b_n$ in this case is almost similar to the case of FRM shown in Fig. \ref{fig:bn_frm}, though,
 there is an initial growth in the LCs for the spin-$1/2$ model which was not present in the FRM case. After the initial growth the 
$b_n$s reach a  peak and then continue down to zero as we reach towards the end of the Krylov chain.

\begin{figure}[h!]
	\begin{minipage}[b]{0.95\linewidth}
		\centering
		\includegraphics[width=0.85\textwidth]{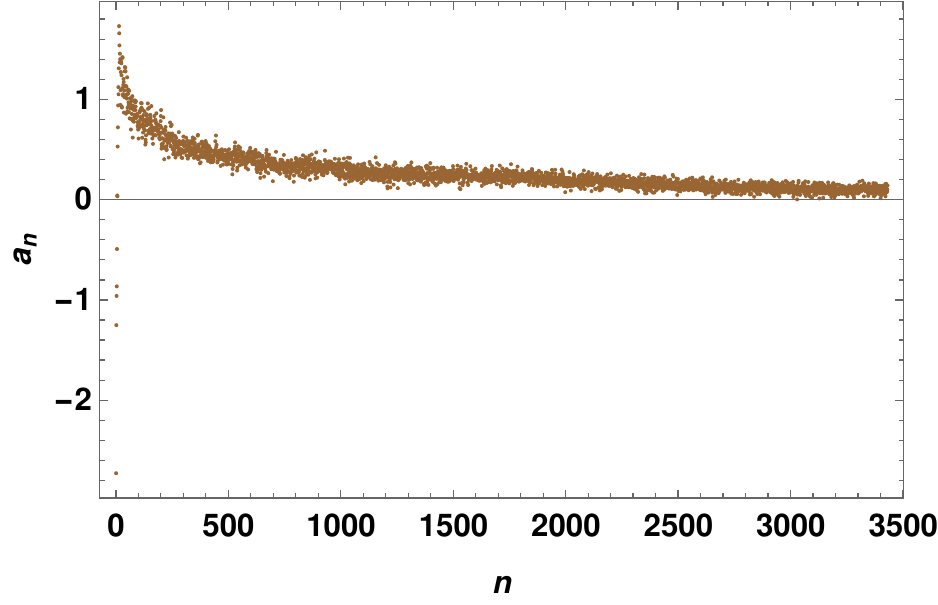}
		\caption{Variation of  $a_n$ with $n$ for a quench in the chaotic limit of the spin-$1/2$ model with 
			disorder.  Here  $L=14$ and the initial state is a domain wall.   }
		\label{fig:an_12cha}
	\end{minipage}
\end{figure}
\begin{figure}
	\begin{minipage}[b]{0.95\linewidth}
		\centering
		\includegraphics[width=0.85\textwidth]{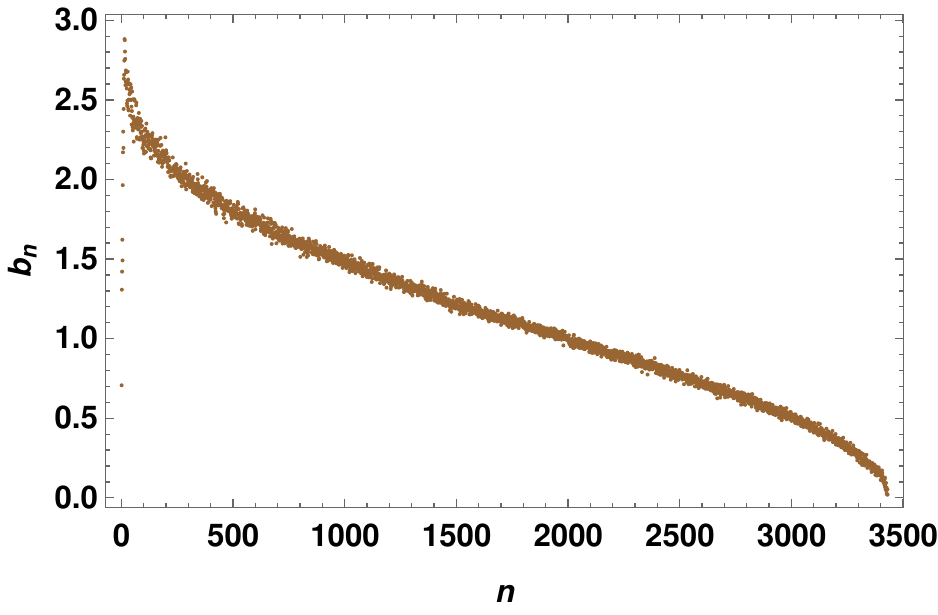}
		\caption{Variation of  $b_n$ with $n$ for a quench in the chaotic spin-$1/2$ model with 
			disorder.  Here  $L=14$ and the initial state is a domain wall.  }
		\label{fig:bn_12cha}
	\end{minipage}
\end{figure}

\begin{figure}[h!]
	\begin{minipage}[b]{0.95\linewidth}
		\centering
		\includegraphics[width=0.85\textwidth]{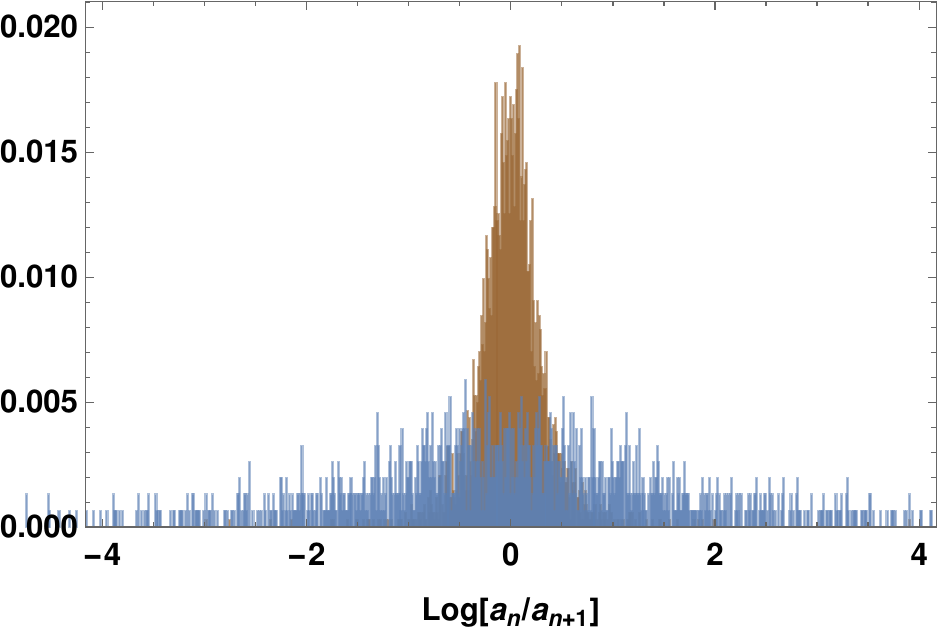}
		\caption{Histogram of the distribution of the LCs $a_n$ for the chaotic (brown) and non-chaotic domains (blue).  }
		\label{fig:hist_an}
	\end{minipage}
\end{figure}

\begin{figure}
	\begin{minipage}[b]{0.95\linewidth}
		\centering
		\includegraphics[width=0.85\textwidth]{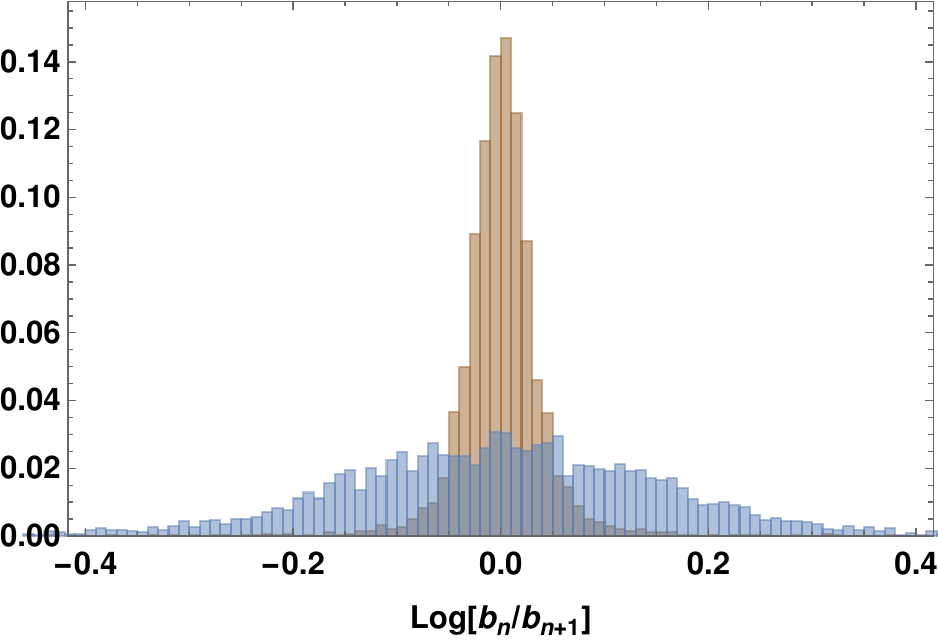}
		\caption{Histogram of the distribution of the LCs $b_n$ for the chaotic (brown) and non-chaotic  (blue) domains.  }
		\label{fig:hist_bn}
	\end{minipage}
\end{figure}

The difference in the distribution of the LCs for the chaotic and non-chaotic domains
can be better visualised from  the histogram   plots for these coefficients shown in Fig. \ref{fig:hist_an} and \ref{fig:hist_bn} for $a_n$ and $b_n$ respectively.  As can be clearly seen from these plots, when the disorder strength is away from the chaotic
domain, the LCs are widely distributed compared to the case when the disorder strength 
is in the chaotic domain. The variance of the distribution of $a_n$ changes from 
$0.584$ in Fig. \ref{fig:an_12int} to a subsequently lower value of $0.449$ corresponding to the chaotic domain in Fig. \ref{fig:an_12cha}. Similarly, variance of the distribution of $b_n$ increases from $0.313$ in the chaotic case (Fig. \ref{fig:bn_12cha}) to $2.319$ (in Fig. \ref{fig:bn_12iint}) as the disorder strength
is increased.


\begin{figure}[h!]
		\centering
		\includegraphics[width=0.45\textwidth]{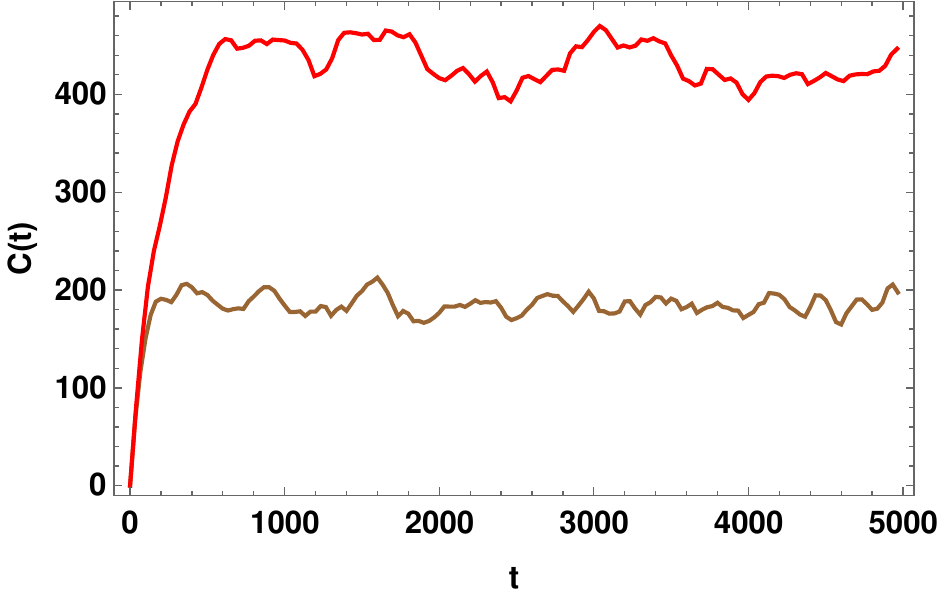}
		\caption{Time evolution of the SC after quench in a chaotic spin-$1/2$ model with disorder, 
			for $L=12$  (brown) and $L=14$ (red).  After a characteristics initial growth,
		the SC attains a saturation at late times. However, the peak in the complexity
		is almost absent in this case.
  }
		\label{fig:sc_12cha}
\end{figure}

The time evolution of the SC in this case is shown in Fig. \ref{fig:sc_12cha} (for reference 
we have also shown the SC with $L=12$ as well). We notice that similar to the case of quench modeled with 
FRM, (shown in fig. \ref{fig:SC_frm1}) there is an initial linear growth at earlier times, however,
the peak in the complexity is less pronounced compared to the FRM case. This 
behaviour of the SC can be understood from the fact that in the case of FRM, the correlation
hole is deeper, while for the disordered spin-1/2 system,  
 as the disorder strength is increased, the correlation hole becomes less pronounced,
and fades for larger value of  disorder.
At late times, the SC fluctuates around the long time average value which
depends on the dimension of the Krylov space, and hence on the system size. Therefore, as 
the strength of the disorder of the quenched spin chain is changed such that the system changes from 
integrable to chaotic, the SC can capture the necessary features 
of the corresponding phase in both the cases.


\section{Summary and Conclusions}\label{summary}

In this paper, we have performed a detailed analysis of the evolution of the SC after quenches in interacting 
quantum many-body systems. We have shown that, for time scales that are small compared to the inverse of the 
width of the LDOS, the SC grows quadratically with time, irrespective of the final Hamiltonian 
 or the 
initial state before the quench, with the rate of the growth being determined by the width of the LDOS (the width, in turn, is equal to the 
LC $b_1$). Behaviour of SC evolution in the next time scale is determined by the initial state, as well as
the strength of the perturbation introduced via the quench. Exponential and Gaussian decays of the SP are two of the most commonly
encountered  behaviours of the  SP after the initial quadratic decay. In the presence of strong perturbations, 
the SP shows Gaussian decay, and hence, the quadratic growth of the SC persists for a time longer than the inverse of the 
width of the LDOS. For sudden quenches in  XX model and spin-$1/2$ systems with impurities, the quadratic growth 
can persists even up to the saturation point of the SP.

On the other hand, when the strength of the external perturbation is not strong, the initial quadratic decay 
merges into an exponential decay at late times. Using an interpolation formula connecting these two different decays,
proposed in \cite{Flambaum}, we have obtained the associated LC and the SC. The LC $b_n$s grows linearly with
$n$ (whereas  $a_n=0$), with the slope of the linear growth being determined by the width of the LDOS,
and the initial quadratic growth of the SC merges into a linear growth at late times due to the 
presence of the exponential decay of the SP.

To understand the behaviour of SC at late times, we first modeled the quenched interacting system as a FRM in the GOE.
The LC and the SC are then obtained by finding out the Hessenberg form of these RMs. Here, $a_n \approx 0$ and the 
$b_n$s can be fitted with a curve of the form $b_n=n_1(\mathcal{N}-n)^{n_2}$, where the two unknown constants have been 
determined by using the exact analytical expression in Eq. \eqref{SP_FRM} available in the literature 
for the SP after a quench with FRM. Due to the presence of the correlation hole
in the SP, the SC grows linearly with time, reaches a peak, after which it saturates to a lower constant value. These features 
of the SC evolution after quench are consistent with the behaviour for the same observed in \cite{Balasubramanian:2022tpr} without
such a quench.

As the next example,
we considered quenches in an interacting spin-$1/2$ model in the presence of nearest neighbour interactions and disorder,
which shows non-integrable behaviour in a particular range of the disorder parameter. For this model, we have obtained the full
sets of LCs and the SC, in both the  chaotic and the intermediate region between the chaotic and many-body localised phases, 
where the initial state before the quench is assumed to be a domain wall.
Though the LCs show similar patterns in this case as with the FRM, the exact details are different. For example, away from the chaotic phase, the LCs are 
distributed randomly, whereas in the non-integrable phase, the distribution of the LCs are more compact.  Importantly, the sequence of 
$b_n$s shows  a linear growth for small values of $n$ and reaches a peak, a behaviour  which is absent for the corresponding $b_n$ sequence
for our FRM analysis. The analysis of the SC shows that in the chaotic phase of this model, it shows a linear 
growth at early times and saturation at late times, with the peak in the intermediate time between them being less pronounced
in this case compared to that of the FRM.

\begin{center}
\bf{Acknowledgements}
\end{center}
The work of TS is supported in part by the USV Chair Professor position at the Indian Institute of Technology, Kanpur.

\appendix

\section{Krylov basis construction and the spread complexity}\label{krylov}

In this appendix we briefly review the  construction of the Krylov basis states using the Lanczos algorithm, 
and the subsequent definition of the 
SC of an arbitrary time evolved initial state after a quench under 
a new Hamiltonian. This procedure has been used in section \ref{early_and_FRM} to 
find out the LC and the SC.

Assume that a sudden  quench is performed on a quantum system at in initial time $t_0=0$, and
subsequently, the state before the quench evolves under the new Hamiltonian $H$.
In the Lanczos algorithm, one constructs new elements of the Krylov basis starting from
an initial one by using the following procedure
\begin{equation}\label{Krylov-basis}
\big| {K_{n+1}} \rangle=\frac{1}{b_{n+1}}\Big[\big(H-a_{n}\big)\big|{K_n}\rangle-b_{n}\big|{K_{n-1}}\rangle\Big]~.
\end{equation}
Here, $\big|{K_{0}}\rangle=\big|{\psi_0}\rangle$, i.e., the first element of the Krylov basis
is the initial state before the quench, and $H$ represents the Hamiltonian after the 
quench. The two sets of coefficients $a_n$ and $b_n$ are the LCs. The $b_n$s
fix the normalization of the Krylov basis vectors at each step of the 
recursion and the $a_n$s are given by the expectation value of the post-quench Hamiltonian
in the Krylov basis, i.e.,
\begin{equation}
a_{n}=\langle {K_{n}|H|K_{n}}\rangle~.
\end{equation}
The recursion stops when $b_n=0$ for some value of $n$.

Now we can expand the time evolved state after quench in terms of the Krylov basis
\begin{equation}
\big|{\psi(t)}\rangle=\sum_{n}\phi_{n}(t)\big|{K_{n}}\rangle~,
\end{equation}
where, using the Schrodinger equation satisfied by the Hamiltonian $H$, one can show that
the expansion coefficients $\phi_{n}(t)$ satisfy the following discrete  Schrodinger equation
\begin{equation}\label{dse}
i\dot{\phi}_{n}(t)=a_{n}\phi_{n}(t)+b_{n}\phi_{n-1}(t)+b_{n+1}\phi_{n+1}(t)~.
\end{equation}  
Here, an overdot represents a derivative with respect to time. In \cite{Balasubramanian:2022tpr} it was proved that if we consider cost functions 
of the form  $\mathcal{C}_{B}(t)=\sum_{n}n|\langle {\psi(t)|B_{n}}\rangle|^2$,
to indicate the spread of a time evolved wavefunction in terms of a complete orthonormal basis
$\big|B_{n}\rangle$, then this cost is minimised when evaluated in the Krylov basis
$\big|K_{n}\rangle$. Therefore, we  arrive at the definition of the SC as the 
minimum of the above cost as
\begin{equation}\label{spread_complexity}
\mathcal{C}(t)=\sum_{n}n|\langle {\psi(t)|K_{n}}\rangle|^2=
\sum_{n}n|\phi_{n}(t)|^2~.
\end{equation}

We conclude this appendix by briefly outlining  the procedure of obtaining the LCs from the 
moments of the return amplitude $\mathcal{S}(t)$ following \cite{Viswanath}. Consider the 
expansion of the  auto-correlation function in terms of the moments 
\begin{equation}\label{moments}
\mathcal{S}(t)=\sum_{n}^{\infty}M_{n}^{*}\frac{t^n}{n!}~.
\end{equation}
The next step is to construct two sets of auxiliary matrices 
$L_{k}^{(n)}$ and $M_{k}^{(n)}$ from the moments $M_{n}^{*}$. The 
two sets of LCs are  then obtained from these auxiliary matrices as 
 $b_{n}=\sqrt{M_{n}^{(n)}}$ and $a_{n}=-L_{n}^{(n)}$,  
where we have to chose  initial conditions properly so that $b_{0}=0$ \cite{Balasubramanian:2022tpr, Viswanath}. Knowing the full set of LCs  we can then solve the discrete Schrodinger equation in \eqref{dse} to obtain the time-dependent expansion 
coefficients $\phi_{n}(t)$, and subsequently obtain the complexity by evaluating the sum 
 in Eq. \eqref{spread_complexity}.


\end{document}